\documentclass[10pt,twocolumn]{article}
\usepackage{graphicx}
\title{Summary of the J-PARC UCN Taskforce}
\author{The J-PARC UCN Taskforce}
\date{Apr. 15, 2009}

\begin{document}
\maketitle

\section{Introduction}
We summarize the discussion on the technical possibility of the realization of ultracold neutron (UCN) source at the Japan Proton Accelerator Research Complex (J-PARC) for new physics beyond the standard model~\cite{ram08,bak06}.
We discussed the UCN production according to the down-scattering in the superfluid helium of very cold neutrons from the spallation reaction with the proton beam from the linear accelerator (LINAC) of the J-PARC.
In this report, we report the summary of taskforce meetings in the period of Jan.-Mar. 2009~\footnote{
2009/01/28 13:00-15:00\\ \hspace{\parindent}
2009/02/04 15:30-16:30\\ \hspace{\parindent}
2009/02/12 16:00-17:00\\ \hspace{\parindent}
2009/02/19 17:00-18:00\\ \hspace{\parindent}
2009/03/11 10:30-16:00}.
In the quickest case, J-PARC can provide UCNs for physics experiments after two-year construction period.
We also expect that the UCN source would be the most intense source in the world at the startup stage and it can be reinforced by one order of magnitude.
We recommend to consolidate the nation-wide and international collaboration to study the details of the source design, experimental techniques and physics experiments.

\section{Distribution of Proton Beam}
\begin{figure*}[htbp]
	\begin{center}
		\includegraphics[width=0.98\linewidth,clip,angle=90]{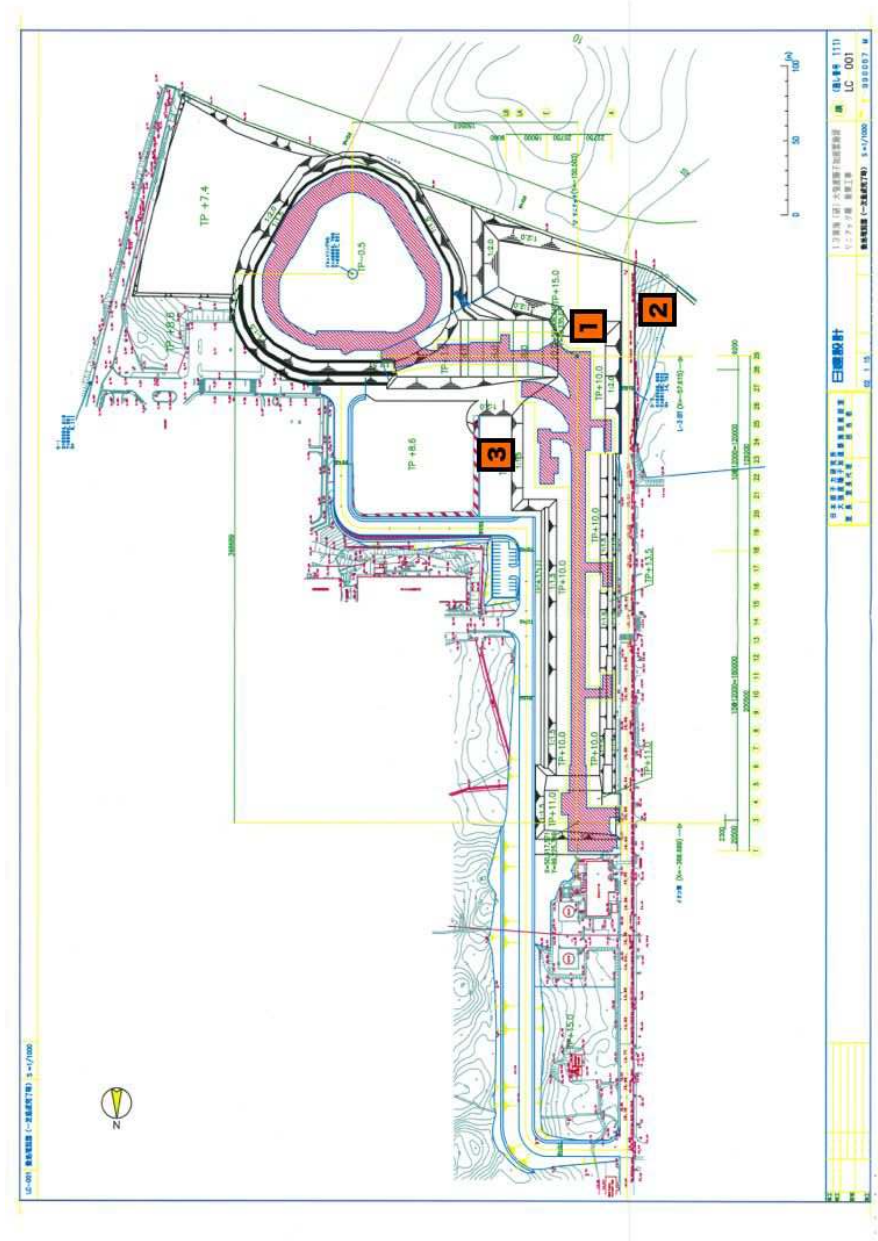}
	\end{center}
	\caption{
	Layout of the LINAC and $3$GeV Rapid Cycle Synchrotron of the J-PARC. 
	1, 2, and 3 in the figure are the UCN source locations discussed in the taskforce.
	}
	\label{fig:locations}
\end{figure*}

The layout of the linear accelerator (LINAC) and the 3 GeV Rapid Cycle Synchrotron (RCS) of the J-PARC is shown in Fig.~\ref{fig:locations}.

\begin{figure*}[htbp]
	\begin{center}
		\includegraphics[width=0.95\linewidth,clip]{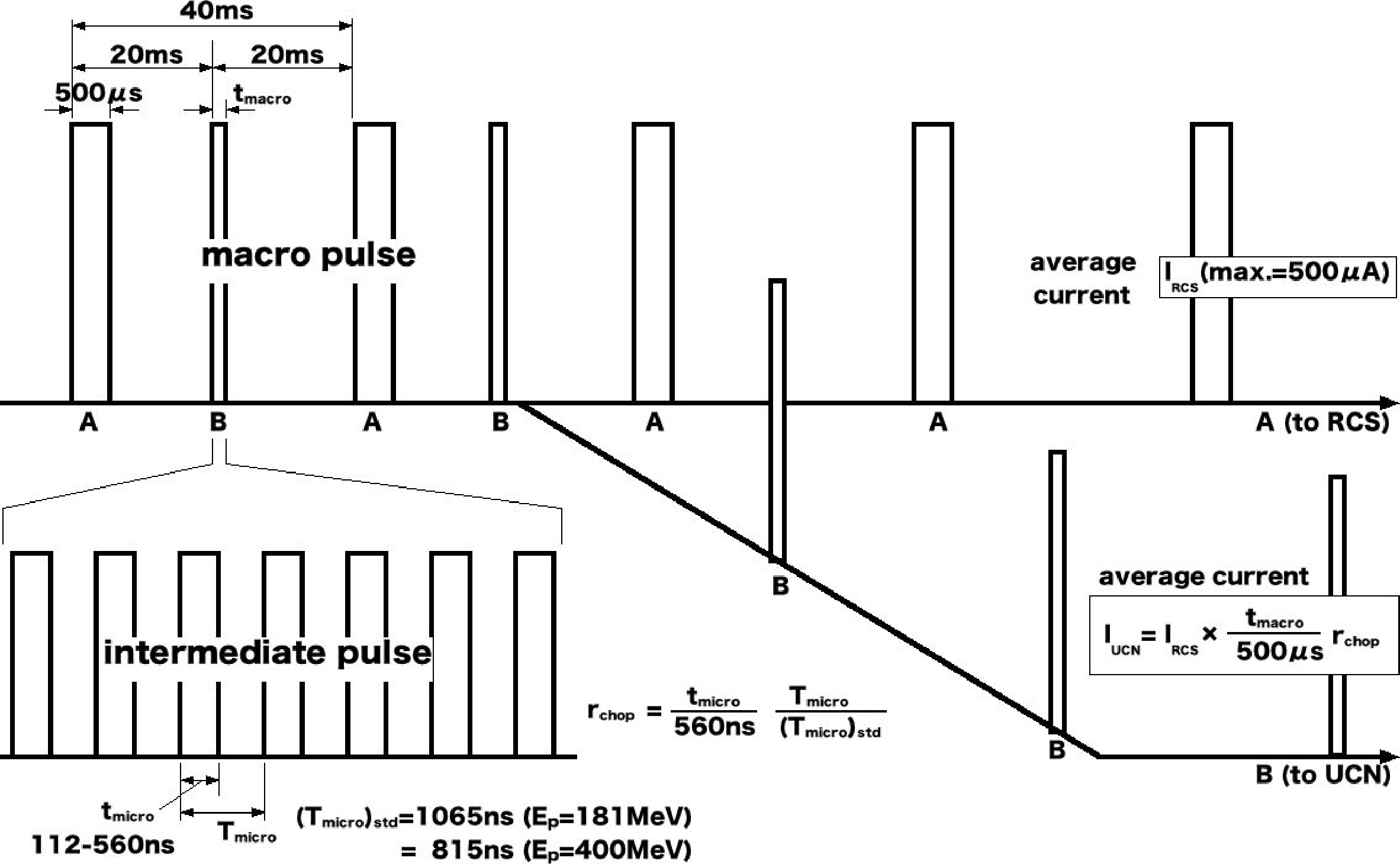}
	\end{center}
	\caption{
		Pattern diagrams of beam distribution of J-PARC LINAC.
	}
	\label{fig:pulse}
\end{figure*}

The J-PARC LINAC is capable of providing pulsed proton beams with the repetition rate of 50 Hz, while it is currently providing protons to the RCS with the repetition rate of 25 Hz.
We consider the case that the beam delivery to the RCS is 25 Hz without disturbing the utilization downstream of the RCS and the additional 25 Hz is delivered to the UCN source as shown in Fig.~\ref{fig:pulse}.
The beam power to the UCN source can be adjusted by combining the following tunable factors.

\begin{enumerate}
\item Macro Pulse Width\\
	The original macro pulse time width is $500\mu$s. 
	The average current value is decreased by shortening this to $t_{\mathrm{macro}}$
	The adjustable range is 
	\begin{equation}
		50 \mu{\mathrm{s}} \le t_{\mathrm{macro}} \le 500 \mu{\mathrm{s}} .
	\end{equation}

\item Chopping Ratio\\
	Each macro pulse is a set of intermediate pulses.
	We denote the width and the repetition rate of the intermediate pulse as $t_{\mathrm{micro}}$ and $T_{\mathrm{micro}}$.
	The standard width is $(t_{\mathrm{micro}})_{\mathrm{std}}=560$ ns.
	The standard repetition rate is chosen for the synchronization to the revolution time of protons in the RCS before the acceleration as
	\begin{equation}
		(T_{\mathrm{micro}})_{\mathrm{std}} = \left\{
			\begin{array}{rl}
				1065 {\mathrm{ns}} & (E_p=181 {\mathrm{MeV}}) , \\
				 815 {\mathrm{ns}} & (E_p=400 {\mathrm{MeV}}) . \\
			\end{array}
			\right.
	\end{equation}
	The average beam current can be changed by adjusting $t_{\mathrm{micro}}$ and $T_{\mathrm{micro}}$.
	We refer to the ratio of the average current to that without chopping as the the chopping ratio. 
	The chopping ratio is given as 
	\begin{equation}
		r_{\mathrm{chop}} = \frac{t_{\mathrm{micro}}}{(t_{\mathrm{micro}})_{\mathrm{std}}} \frac{T_{\mathrm{micro}}}{(T_{\mathrm{micro}})_{\mathrm{std}}} .
	\end{equation}
	The chopping ratio can be lowered by shortening the width and lowering the repetition rate of the intermediate pulse.
	The pulse width in the range of
	\begin{equation}
		\frac{1}{5} \le \frac{t_{\mathrm{micro}}}{(t_{\mathrm{micro}})_{\mathrm{std}}} \le 1 
	\end{equation}
	is already proven experimentally.
	On the other hand, the adjustable range of the repetition rate with the present control system is
	\begin{equation}
		\frac{T_{\mathrm{micro}}}{(T_{\mathrm{micro}})_{\mathrm{std}}}=1, \hspace{3pt} \frac{1}{2}, \hspace{3pt} \frac{1}{4} .
	\end{equation}
	Therefore, the proven range in the chopping ratio is 
	\begin{equation}
		0.05 \le r_{\mathrm{chop}} \le 1 .
	\end{equation}
\item Switch On/Off of Macropulses\\
	The average beam current can be decreased by switching off the beam supply from the ioninzation source.
	The macropulse rate fraction to the UCN source can be adjusted in the range of 
	\begin{equation}
		0 \le \frac{N_{\mathrm{UCN}}}{25 {\mathrm{Hz}}} \le 1 ,
	\end{equation}
	where we denote the macropulse rate to the UCN source as $N_{\mathrm{UCN}}$
\end{enumerate}

The peak current ($I_{\mathrm{peak}}$) is presently $I_{\mathrm{peak}}=30$mA with $E_p=181$MeV, which is to be upgraded to $I_{\mathrm{peak}}=50$mA with $E_p=400$MeV. 
We expect that the average current ($I_{\mathrm{UCN}}$) to the UCN source can be adjusted in the range of 
\begin{eqnarray}
0 \le I_{\mathrm{UCN}} &\le& 
	\frac{t}{500 \mu{\mathrm{s}}} r_{\mathrm{chop}} \frac{N_{\mathrm{UCN}}}{25 {\mathrm{Hz}}} \nonumber\\
	&\times& \left\{
	\begin{array}{ll}
		375 \mu{\mathrm{A}} & (E_p=181, 400{\mathrm{MeV}})\\
		625 \mu{\mathrm{A}} & (E_p=400{\mathrm{MeV}})
	\end{array}
	\right.
\end{eqnarray}
together with the arbitrary macropulse pattern.
The available proton beam power to the UCN source $W_{\mathrm{UCN}}$ is in the following range.
\begin{equation}
0 \le W_{\mathrm{UCN}} \le
	\left\{
	\begin{array}{ll}
		67.9 {\mathrm{kW}} & (E_p=181{\mathrm{MeV}}) \\
		150 {\mathrm{kW}} & (E_p=400{\mathrm{MeV}}) \\
		250 {\mathrm{kW}} & (E_p=400{\mathrm{MeV}})
	\end{array}
	\right.
	\label{eq:power}
\end{equation}
In this report, we take the case that $t_{\mathrm{PW}}=50$ and $N_{\mathrm{UCN}}=25$ Hz, which leads to 
\begin{equation}
W_{\mathrm{UCN}} = 20 {\mathrm{kW}} .
\end{equation}

\section{Proton Beam Transport}
The feasibility to transport the proton beam to the candidates of UCN source locations location of the UCN source has been discussed under following restrictions.
\begin{itemize}
\item Curvature Radius of Beam Orbit\\
The output of the LINAC is H$^-$ ions.
The strength of the magnetic field to bend the H$^-$ beam is restricted to be less than $0.5$T to avoid the strip-off of the electrons.

An alternative solution is to convert the charge state from H$^-$ to H$^+$ by installing a charge stripping foil, which suppresses the orbit curvature radius by a factor of 2.
However, we assume to use the H$^-$ beam to avoid additional efforts to maintain the stripping foil and the radiation shields.

\item Beam Size\\
The lifetime of the proton incident window is inversely proportional to the proton current density as Eq.~\ref{eq:window-life}.
The proton current density needs to be appropriately controlled for a reasonable operation lifetime since the proton incident window of the neutron production target experiences the most severe radiation damage.
We require that the beam size is expanded to be about $10 \sim 25$mm (r.m.s.) on the proton incident window.
\end{itemize}

\subsection{UCN Location Candidate 1: Extension of the LINAC downstream to the south}
\begin{figure*}[htbp]
	\begin{center}
		\includegraphics[width=0.95\linewidth,clip]{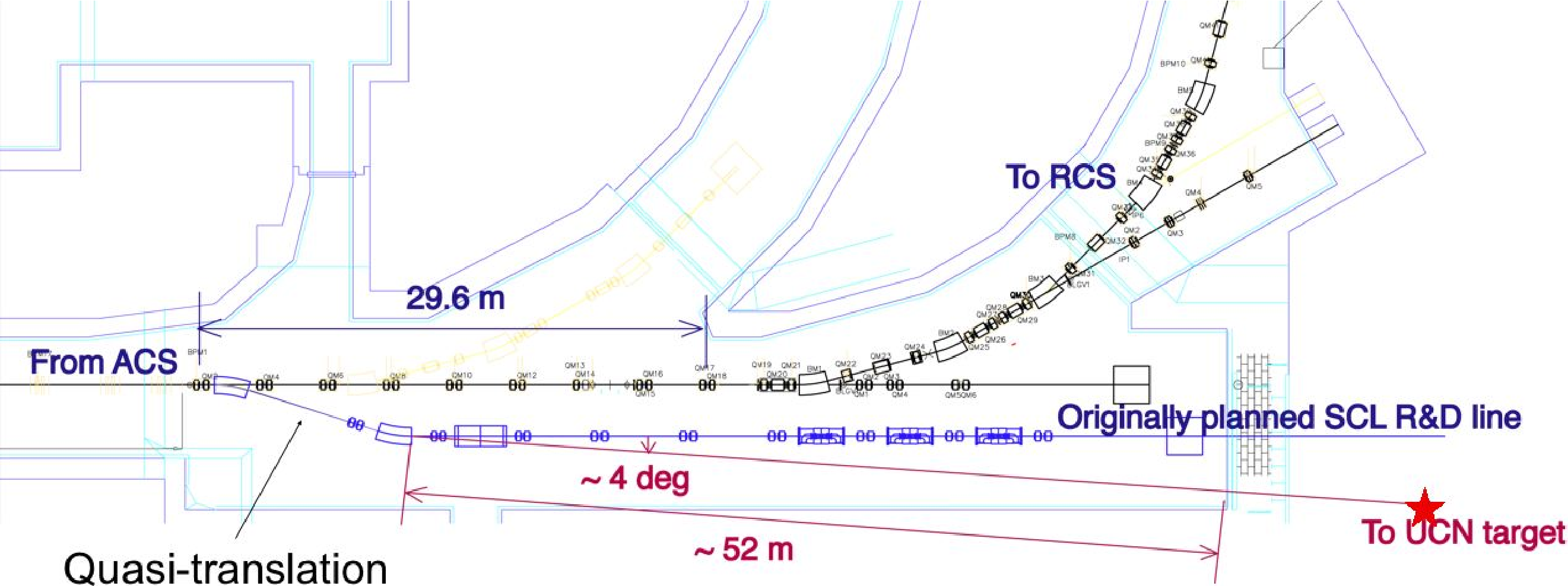}
	\end{center}
	\caption{
		UCN source location candidate 1.
	}
	\label{fig:branch-1}
\end{figure*}

Here we discuss the case of the UCN source is located at the position $1$ in the Fig.~\ref{fig:locations}.
The proton beam is branched as shown in Fig.~\ref{fig:branch-1}.
The proton beam is kicked off to the west and transported off the original LINAC line so that it does not disturb the "Originally planned SCL R\&D line" of Fig.~\ref{fig:branch-1}, which is temporarily reserved for the future extension to the 2nd neutron target station.
We assume that the UCN source would be installed in the extended building on the same level as the floor level of the LINAC building.

\begin{figure*}[htbp]
	\begin{center}
		\includegraphics[width=0.9\linewidth,clip]{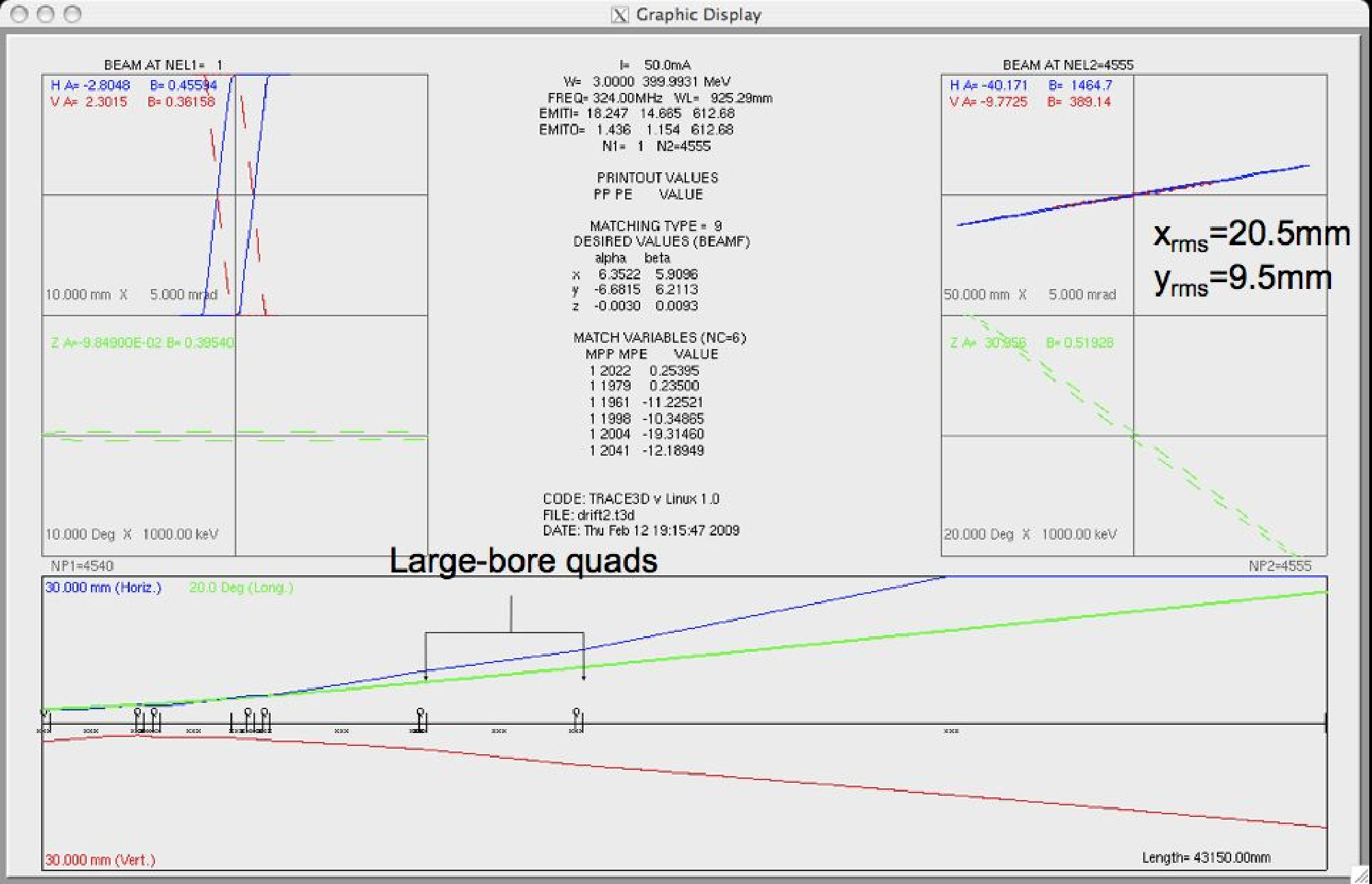}
	\end{center}
	\caption{
		Proton beam optics to the UCN source location candidate 1. 
	}
	\label{fig:branch-1-opt}
\end{figure*}

\begin{figure*}[htbp]
	\begin{center}
		\includegraphics[width=0.5\linewidth,clip]{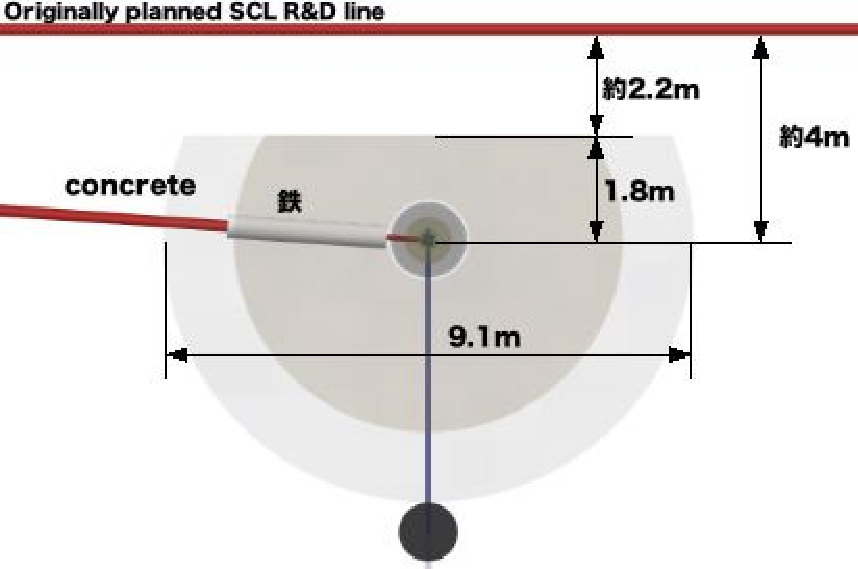}
	\end{center}
	\caption{
		Relation between the UCN source and the "Originally planned SCL R\&D line".
	}
	\label{fig:line-relation}
\end{figure*}

The proton beam is transported for about $60$ m with the angle of 4${\circ}$ off the LINAC axis.
The proton beam optics is shown in Fig.~\ref{fig:branch-1-opt}. 
The beam size at the target position is $x_{\mathrm{rms}} \simeq 20$mm and $y_{\mathrm{rms}} \simeq 10$mm.
The window lifetime is estimated to be about 27 years according to Eq.~\ref{eq:window-life} with the operation time of 5000 hours per year even for $E_p=181$MeV case, which is long enough in terms of the maintenance efforts.
The relation between the UCN source and the "Originally planned SCL R\&D line" is shown in Fig.~\ref{fig:line-relation}.
The width for the LINAC extension is restricted in about 10 m long region.
The size of the neutron source is described later (c.f. section \ref{sec:shield}). 

\subsection{UCN Location candidate 2: Extension of the LINAC to southwest}
\begin{figure*}[htbp]
	\begin{center}
		\includegraphics[width=0.95\linewidth,clip]{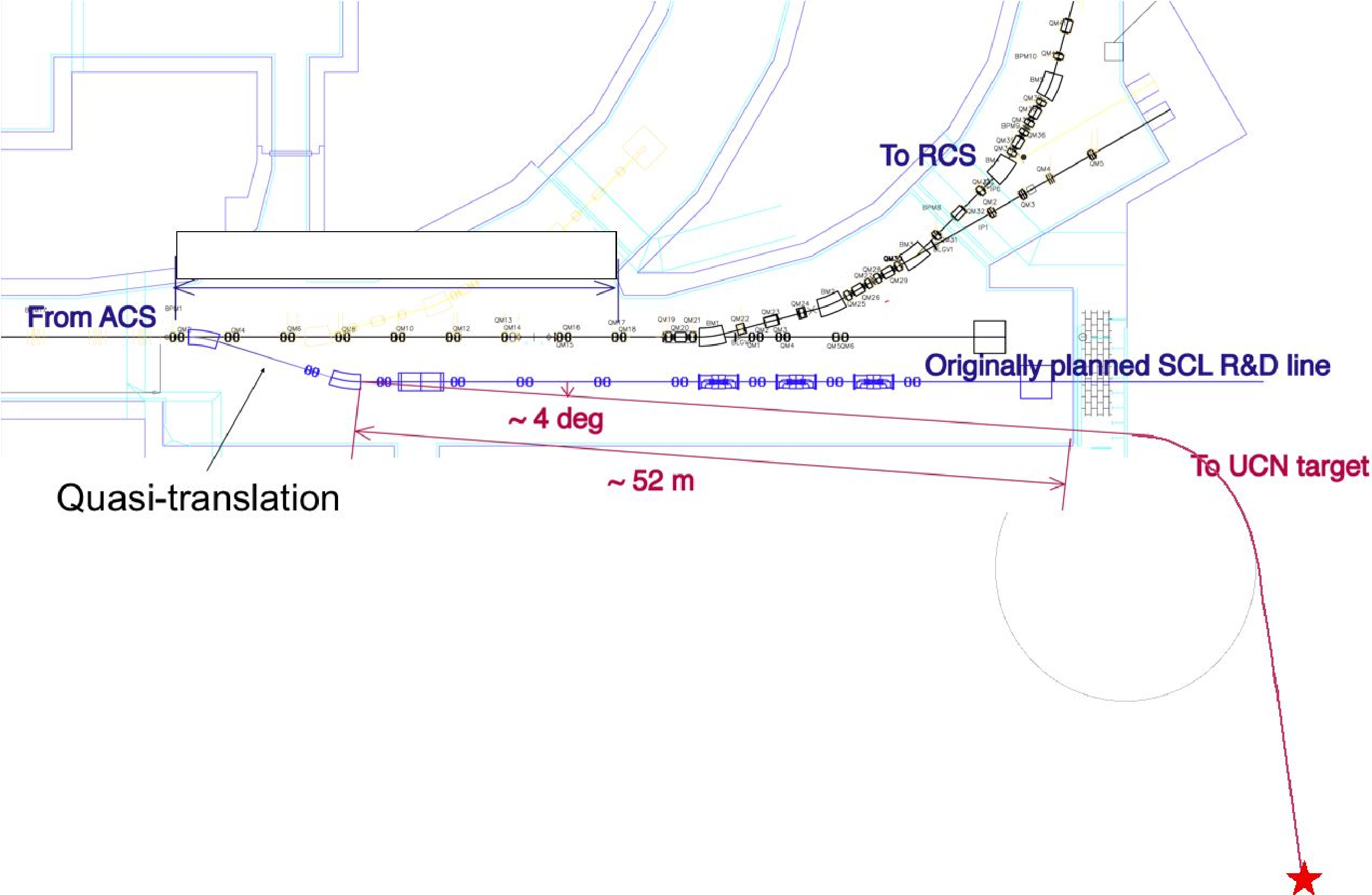}
	\end{center}
	\caption{
		UCN source location candidate 2. 
	}
	\label{fig:branch-2}
\end{figure*}

It is possible to transport the proton beam further off the LINAC axis to the location $2$ in Fig.~\ref{fig:locations} by additionally bending to the west as shown in Fig.~\ref{fig:branch-2}.
However, this candidate has not been discussed in details.

\subsection{Location Candidate 3: U-turn Layout}
The location $3$ in the Fig.~\ref{fig:locations} is also possible by bending the proton beam by 180$^{\circ}$.

In this case, the first $90^{\circ}$ bending has the same curvature as the L3BT transport line in the existing tunnel.
The beam is branched from the ADS line and bent by another $90^{\circ}$.
The beam defocus is initiated after the beam branch from the ADS line so that it does not affect the beam optics to the ADS line.

The quadrupole magnets are not installed in the second $90^{\circ}$ bending for minimizing the curvature radius since the achromat beam transport is not necessary for the UCN line.
The curvature radius of $8.3$ m can be realized by employing six magnets of $1.67$ m long, $15^{\circ}$ bending angle with $6.37$ m curvature radius placed with the spacing of $0.6$ m.

However, the drift space for expansion of the beam size can be decreased by installing two or three quadrupole magnets during the $90^{\circ}$ bending.
The length of the drift space is estimated as $20$ m downstream of the $90^{\circ}$ bending.

\begin{figure*}[htbp]
	\begin{center}
		\includegraphics[width=0.95\linewidth,clip]{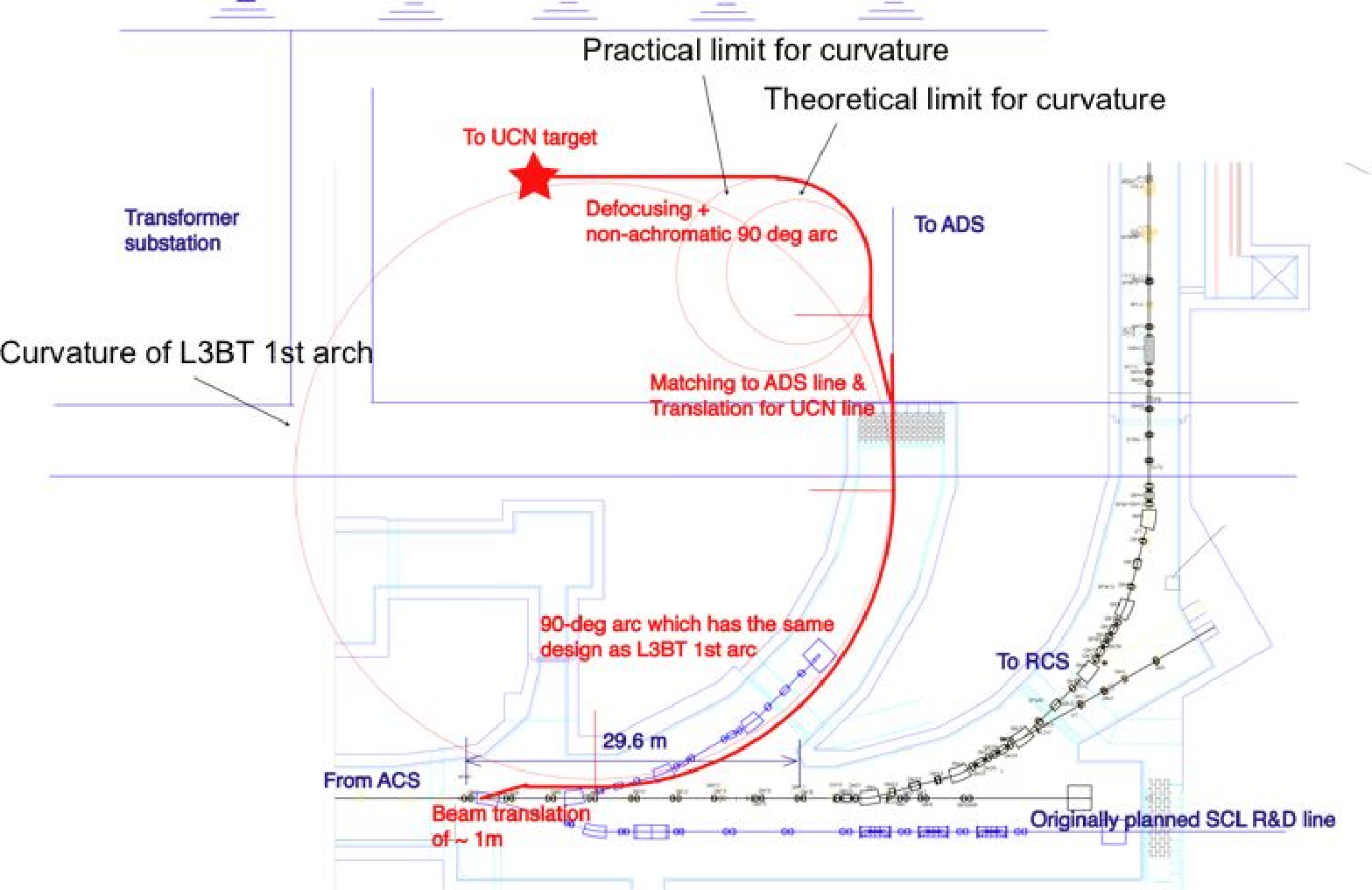}
	\end{center}
	\caption{
		UCN source location candidate 3. 
	}
	\label{fig:branch-3}
\end{figure*}

The corresponding UCN source location is shown in Fig.~\ref{fig:branch-3}~\footnote{
If we use H$^+$ beam, the curvature radius can be half since the magnetic field can be twice.
However, the drift space required for the beam defocus doesn't change and only the curvature of the second $90^{\circ}$ bending can be shortened.
We adopt H$^-$ beam transport considering both of the advantage and disadvantage.
}.
In this design, the beam size on the neutron production target is $2x_{\mathrm{rms}} \sim 15$mm and $2y_{\mathrm{rms}} \sim 15$mm.
The beam size corresponds to the window lifetime of about $25$ years, which is reasonably long. 

\begin{figure*}[htbp]
	\begin{center}
		\includegraphics[width=0.95\linewidth,clip]{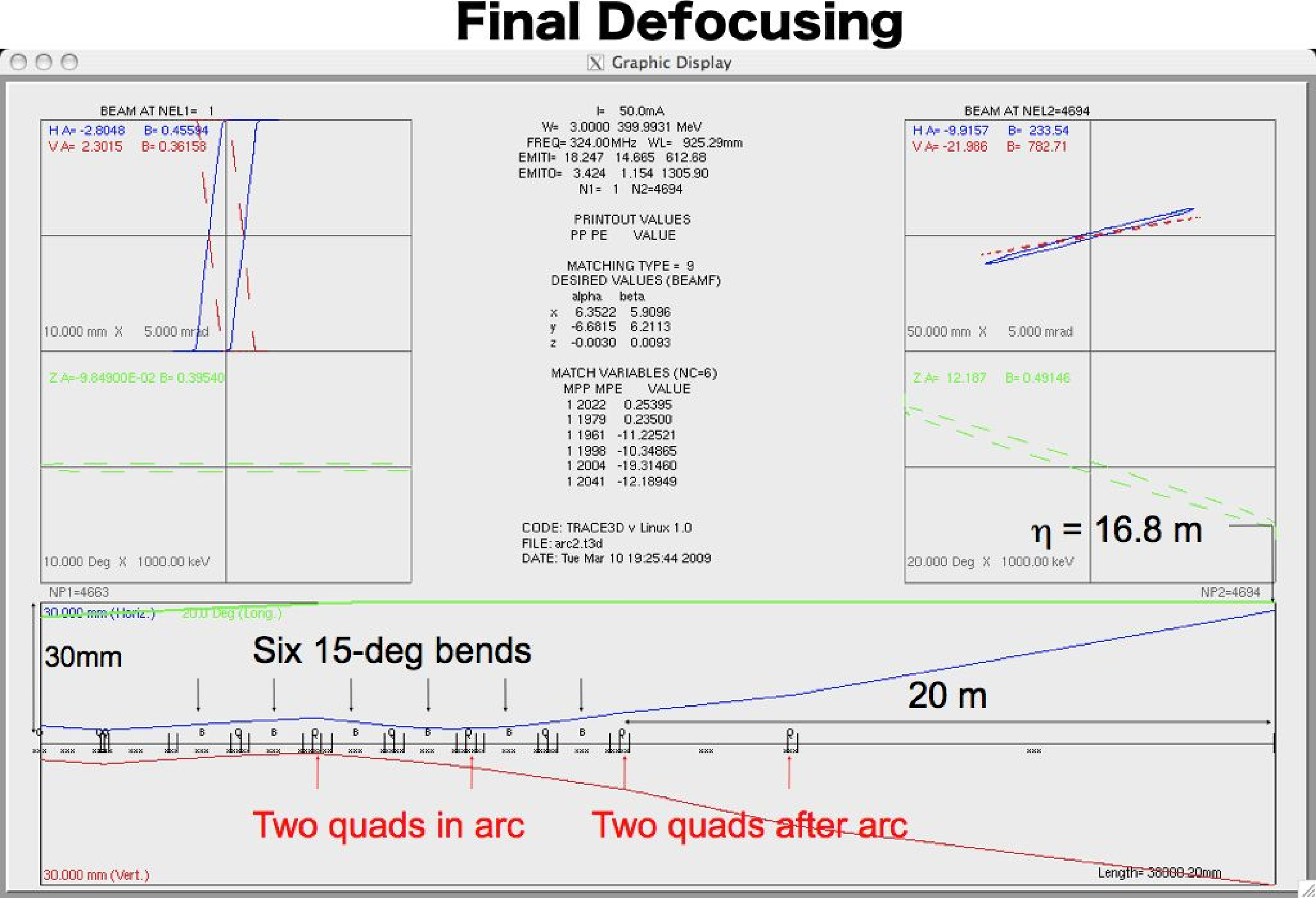}
	\end{center}
	\caption{
		Proton beam optics to UCN source location candidate 3. 
		$2x_{\mathrm{rms}}$ and $2y_{\mathrm{rms}}$ are shown.
	}
	\label{fig:branch-3-opt}
\end{figure*}

\section{Ultracold Neutron Source}
We assume the proton beam with the kinetic energy of $181$ MeV or $400$ MeV is incident to the neutron production target with the beam power of $20$ kW.
Neutrons are generated by the spallation reaction and moderated into the cold region.
The cold neutrons are converted to ultracold neutrons according to the inelastic scattering in superfluid helium.
We also assume that ultracold neutrons are accumulated in the superfluid helium.

\begin{table}
	\begin{center}
	\begin{tabular}{|c|c|}
	\hline
	$E_p$ & $(I_{\mathrm{UCN}})_{\mathrm{max}}$ \\
	proton kinetic energy & beam current\\
	& for $20$ kW operation\\
	\hline
	$181$ MeV & $110 \mu$A \\
	$400$ MeV & $50 \mu$A \\
	\hline
	\end{tabular}
	\caption{
	Beam current for $20$kW beam power. 
	}
	\label{tab:Imax}
	\end{center}
\end{table}

\subsection{Production Target}
\begin{figure}[htbp]
	\begin{center}
		\includegraphics[width=0.95\linewidth,clip]{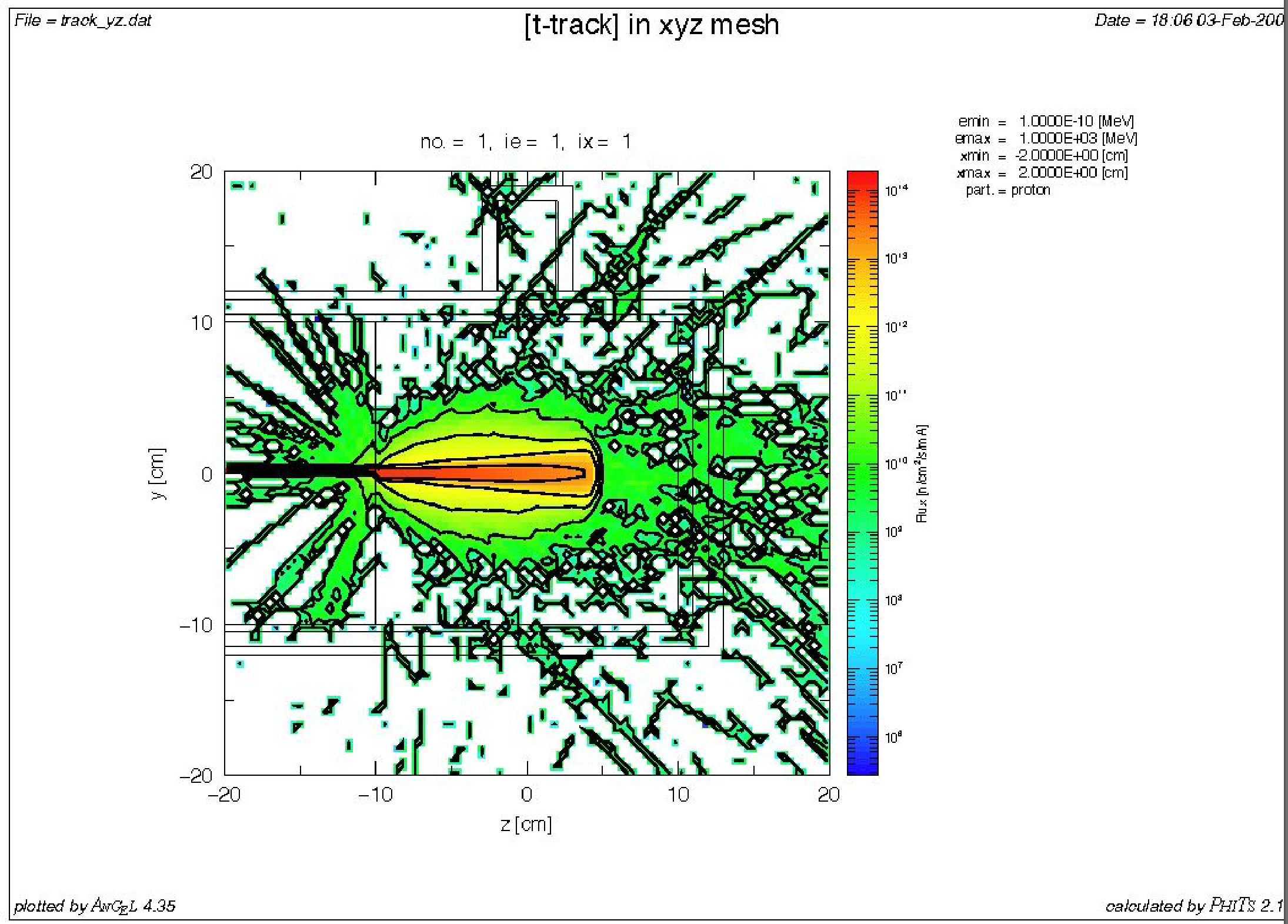}
	\end{center}
	\caption{
		Distribution of $400$ MeV protons in the Pb target. 
	}
	\label{fig:target}
\end{figure}

The range of a $400$ MeV proton in Pb is about $13$ cm as shown in Fig.~\ref{fig:target} (simulation code is PHITS~\cite{phits}). 
We, hereafter, adopt the use of a $20$ cm thick Ta-coated W target, which has been established as the spallation neutron target material. 

The proton incident window of the neutron production target experiences the most severe radiation damage.
The lifetime of the window is estimated by evaluating the displacement per atom (dpa)~\footnote{
Although an additional effect of brittleness caused by He and H from the nuclear reaction induced by $3$ GeV proton beam, the dpa evaluation is considered appropriate for $400$ MeV protons.
}.

We estimate the case of aluminum alloy A5083 which is used in the J-PARC MLF neutron source.
The dpa rate is estimated as $\mathrm{dpa}=0.44$ y$^{-1}$ for the peak current density of $4.3\mu$A cm$^{-2}$ with the $5000$ hour operation in a year~\cite{har05}.
If we assume the acceptable dpa is $({\mathrm{dpa}})_{\mathrm{max}}=10$, the lifetime is given as $t_{E_p=3{\mathrm{GeV}}}^{\mathrm{dpa}=10} \ge 22.7$ y~\footnote{
There is another estimation with the peak current of $10 \mu$A cm$^{-2}$ and $({\mathrm{dpa}})_{\mathrm{max}}=5$, which leads to $t_{E_p=3{\mathrm{GeV}}}^{\mathrm{dpa}=5} \ge 11.3$ y.
}.

One of the difference between our case and J-PARC MLF neutron source is the proton energy.
The Fig.2 in the reference \cite{har05} shows the displacement cross section $\sigma_d$ at $E_p=181$ MeV, $400$ MeV and $3$ GeV are crudely related as
\begin{eqnarray}
\sigma_d(E_p=181[{\mathrm{MeV}}]) &\sim& \sigma_d(E_p=400[{\mathrm{MeV}}]) \nonumber\\
&\sim& \sigma_d(E_p=3[{\mathrm{GeV}}]) \times 1.3 . \nonumber\\ &&
\end{eqnarray}

Another difference is the beam distribution.
While the proton beam position is scanned to paint the window area so that the average beam density becomes as uniform as possible for the J-PARC MLF neutron source, the beam distribution can be approximated by a Gaussian distribution for our case.
We assume the root-mean-square along $x$-axis (horizontal) and that along $y$-axis (vertical) as $x_{\mathrm{rms}}$ and $y_{\mathrm{rms}}$.
We put the average current density in the region of $-2x_{\mathrm{rms}} \le x \le 2x_{\mathrm{rms}}$ and $-2y_{\mathrm{rms}} \le y \le 2y_{\mathrm{rms}}$ as $j_0$.
The current density at the beam center is given as $j_{\mathrm{c}} \simeq 2.8 j_0$ and can be written as
\begin{equation}
j_c \simeq 2.8 \frac{I_{\mathrm{ave}}}{16 x_{\mathrm{rms}} y_{\mathrm{rms}}} ,
\end{equation}
where $I_{\mathrm{ave}}$ is the average current.
The average current is $I_{\mathrm{ave}} =1.1\times 10^{-4}$ A and $I_{\mathrm{ave}} =5\times 10^{-5}$ A for $E_p=181$MeV and $E_p=400$MeV if we assume the average beam power of $W_{\mathrm{UCN}} =20$ kW.

Therefore, we can estimate the window lifetime as
\begin{eqnarray}
t_{E_p=181 {\mathrm{MeV}}}^{\mathrm{dpa}=10} &=& t_{E_p=3 {\mathrm{GeV}}}^{\mathrm{dpa}=10} 
\frac{\sigma_d(E_p=3[{\mathrm{GeV}}])}{\sigma_d(E_p=181[{\mathrm{MeV}}])} \nonumber\\&&\times
\frac{4.3 \mu{\mathrm{A}} {\mathrm{cm}}^{-2}}{2.8 I_{\mathrm{ave}}/(16 x_{\mathrm{rms}} y_{\mathrm{rms}})} \nonumber\\
&=& \frac{x_{\mathrm{rms}}}{10 {\mathrm{mm}}} \frac{y_{\mathrm{rms}}}{10 {\mathrm{mm}}} \times 11 {\mathrm{year}} \nonumber\\
&=& \frac{x_{\mathrm{rms}}}{25 {\mathrm{mm}}} \frac{y_{\mathrm{rms}}}{25 {\mathrm{mm}}} \times 68 {\mathrm{year}} \\
t_{E_p=400 {\mathrm{MeV}}}^{\mathrm{dpa}=10} &=& t_{E_p=3 {\mathrm{GeV}}}^{\mathrm{dpa}=10} 
\frac{\sigma_d(E_p=3[{\mathrm{GeV}}])}{\sigma_d(E_p=400[{\mathrm{MeV}}])} \nonumber\\&&\times
\frac{4.3 \mu{\mathrm{A}} {\mathrm{cm}}^{-2}}{2.8 I_{\mathrm{ave}}/(16 \sigma_x \sigma_y)} \nonumber\\
&=& \frac{x_{\mathrm{rms}}}{10 {\mathrm{mm}}} \frac{y_{\mathrm{rms}}}{10 {\mathrm{mm}}} \times 24 {\mathrm{year}} \nonumber\\
&=& \frac{x_{\mathrm{rms}}}{25 {\mathrm{mm}}} \frac{y_{\mathrm{rms}}}{25 {\mathrm{mm}}} \times 150 {\mathrm{year}}
\label{eq:window-life}
\end{eqnarray}

\subsection{Neutron Moderator}
We consider the conversion of cold neutrons into ultracold region ($E_{\mathrm{UCN}} =252$ neV) out of thermal equilibrium according to the inelastic scatteing.
An intense radiation field of neutrons as cold as possible is necessary to increase the ultracold neutron intensity.
A $20$ kW cold source provides the neutron flux of about $\Phi_0 \simeq 2 \times 10^{12} $cm$^{-2} $s$^{-1}$, which is crudely estimated assuming the $400$ MeV protons are incident to a Pb target with the average current of $50 \mu$A and moderated in D$_2$ with the reflector of Be and graphite as shown in Fig~\ref{fig:tmras-20kW}.
THe neutron temperature is about $30$K. 

\begin{figure}[htbp]
	\begin{center}
		\includegraphics[width=0.8\linewidth,clip]{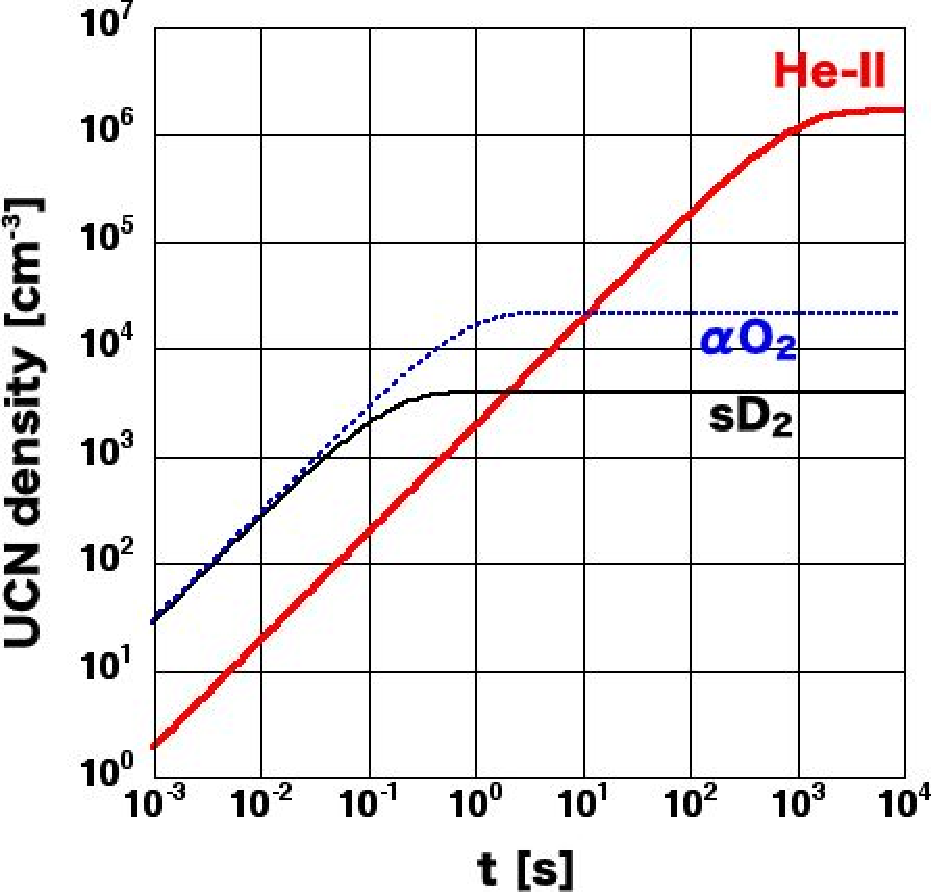}
	\end{center}
	\caption{
Time evolution of the density of ultracold neutrons accumulated in the converter assuming that the temperature of cold neutrons is $30$ K (Boltzmann distribution) and the flux is $\Phi_0=2 \times 10^{12} $ cm$^{-2} $ s$^{-1}$.
	}
	\label{fig:ucn-density}
\end{figure}

The ultracold neutron density accumulated in the converter can be given as
\begin{equation}
\rho_{\mathrm{UCN}} = P \tau (1-e^{-t/\tau}) ,
\label{eq:density-evolution}
\end{equation}
where $P$ is the conversion rate, $\tau$ storage time.
The values of $P$ and $\tau$ are listed in table~\ref{tab:UCN}~\cite{liu02}.
The time evolution of Eq.~\ref{eq:density-evolution} using the values in table~\ref{tab:UCN} is shown in Fig.~\ref{fig:ucn-density}.
The UCN density strongly depends on the volume ratio and geometrical arrangement of converter volume and storage volume.
However, in this summary, we assume that the converter volume is the same as the storage volume.
Therefore, the value is correct as it is, and remaining corrections are the UCN loss on the storage wall, due to upscattering and absorption in the converter.

\begin{table*}
	\begin{center}
	\begin{tabular}{|l|c|c|c|}
	\hline
	converter & He-II & Solid ortho-D$_2$ & $\alpha$-O$_2$ \\
	\hline
	excitation & phonon & phonon & magnon \\
	\hline
	converter temperature & 0.7 K & 5 K & 2 K \\
	\hline
	appropriate neutron temperature & 9 K & 29 K & 12 K \\
	\hline
	conversion rate & & & \\
	$\frac{P}{1{\mathrm{cm}}^{-3}{\mathrm{s}}^{-1}} \frac{2 \times 10^{12}{\mathrm{cm}}^{-2}{\mathrm{s}}^{-1}}{\Phi_0}$ & 1,900 & 26,000 & 28,000 \\
	@(cold neutron temperature of 30K) & & & \\
	\hline
	storage time $\tau$ & 886 s & 0.146 s & 0.750 s \\
	@(wall loss and up-scatteirng ignored) & & & \\
	\hline
	maximum UCN density in the converter & 1,600,000 & 3,800 & 21,000 \\
	$\frac{\rho_{\mathrm{UCN}}}{1{\mathrm{cm}}^{-3}} \frac{2 \times 10^{12}{\mathrm{cm}}^{-2}{\mathrm{s}}^{-1}}{\Phi_0}$ & & & \\
	\hline
	\end{tabular}
	\caption{
Characteristics of UCN converter materials~\cite{liu02}.
The UCN density strongly depends on the volume ratio and geometrical arrangement of converter volume and storage volume.
However, in this summary, we assume that the converter volume is the same as the storage volume.
Therefore, the value is correct as it is, and remaining corrections are the UCN loss on the storage wall, due to upscattering and absorption in the converter.
	}
	\label{tab:UCN}
	\end{center}
\end{table*}

\begin{figure}[htbp]
	\begin{center}
		\includegraphics[width=0.8\linewidth,clip]{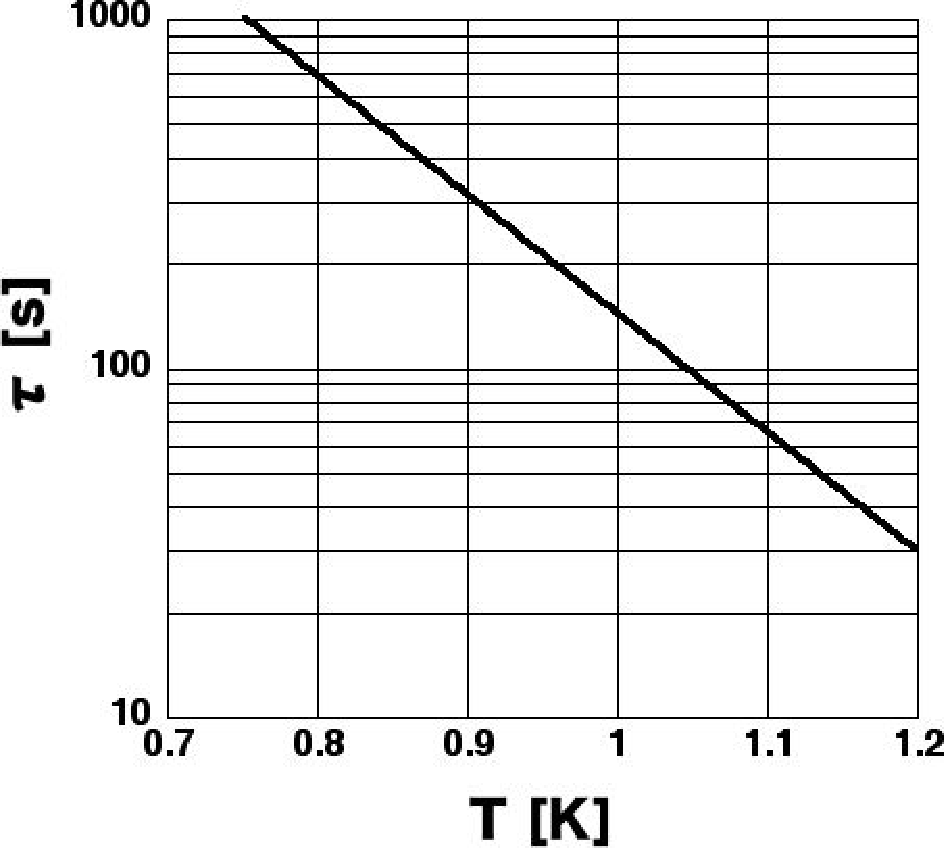}
	\end{center}
	\caption{
		Temperature dependence of the time constant of upscattering in the superfluid helium (He-II). 
	}
	\label{fig:phonon-tau}
\end{figure}

The reversal process of the super-cold neutron generation happens in the superfluidity helium, too. 
It happens by short odds by calling scattering (upscatteirng) on, and the phonon density high (By the temperature of the superfluiditive helium high) as for the reversal process. 
As for the constant, when a super-cold neutron is lost by the above one scattering, what shown as a function of the temperature of the superfluiditive helium is Fig.~\ref{fig:phonon-tau}. 
Because the time constant of the loss by the above one scattering is a neutron longevity level if the temperature of the superfluiditive helium is kept about less than 0.8K, the loss by the above one scattering can be disregarded~\footnote{
Because the neutron absorption by $^3$He that exists naturally by about 1ppm cannot be disregarded when actually accumulating in the superfluiditive helium, it is necessary to use super-high purity (concentrate $^4$He). 
}.

\subsection{Heat Load}
\begin{figure}[htbp]
	\begin{center}
		\includegraphics[width=0.8\linewidth,clip]{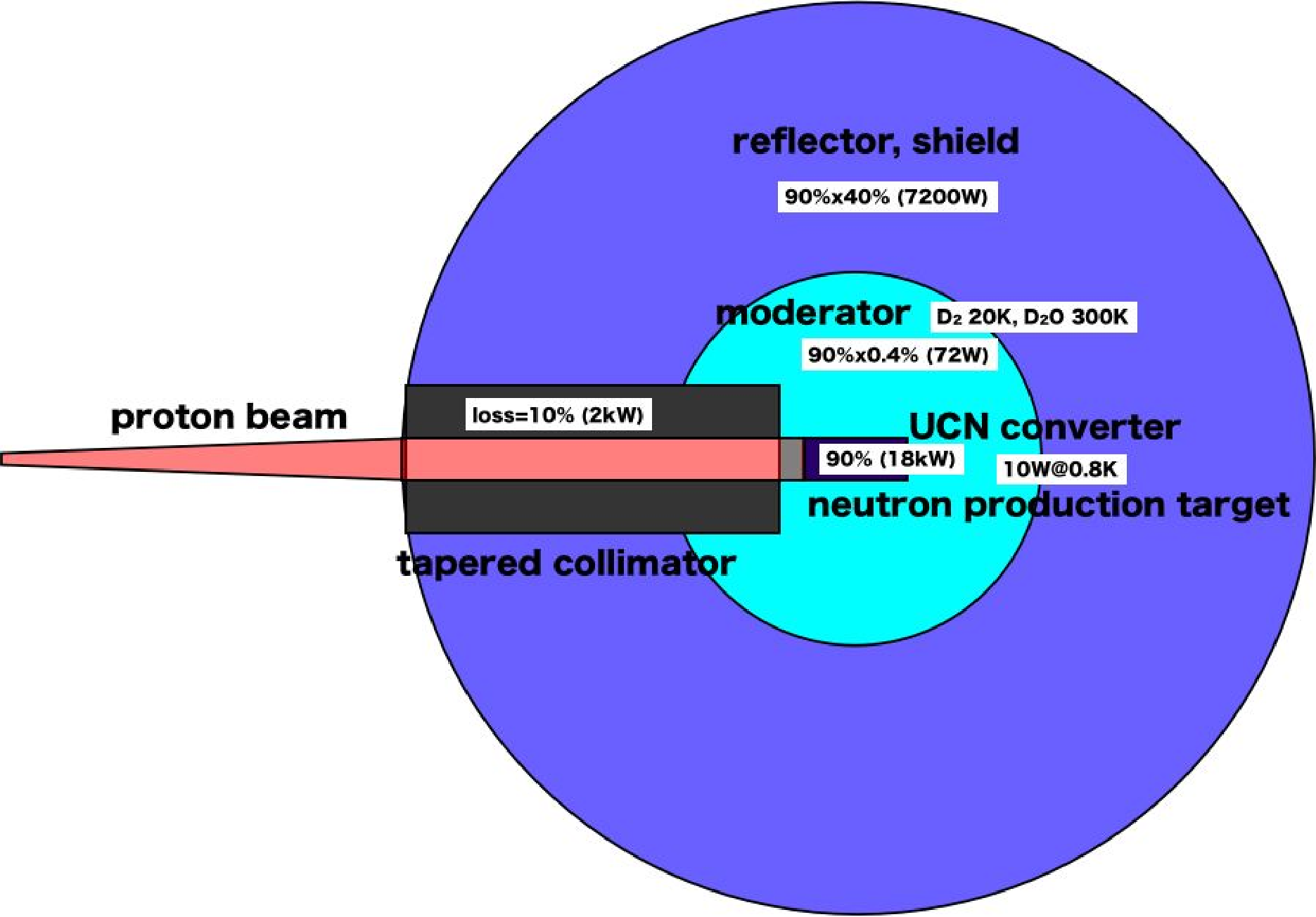}
	\end{center}
	\caption{
		Distribution of the heat load.
	}
	\label{fig:thermal}
\end{figure}

The incident beam would be tailored by a tapered collimator so that the central 90\% beam power reaches to the neutron production target.
The neutron production target is the Ta-coated W.
The target thickness is $20$ cm and the cross section is almost as large as the beam cross section.
About 10\% of the beam power is consumed in the spallation reaction.
The distribution of the heat load is estimated following the design of the J-PARC MLF neutron source as shown in table~\ref{tab:thermal} and Fig.~\ref{fig:thermal}.
\begin{table}
	\begin{center}
	\begin{tabular}{|c|c|c|}
	\hline
	 & heat load [W] & fraction \\
	\hline
	collimator & 2000 & 0.1 \\
	neutron production target & 9000 & 0.45 \\
	moderator & 72 & 0.0036 \\
	UCN converter & 10 & - \\
	reflector and shield & 7200 & 0.36 \\
	\hline
	\end{tabular}
	\caption{
Distribution of the heat load.
Missing heat is consumed in the spallation reaction.
	}
	\label{tab:thermal}
	\end{center}
\end{table}

The heat load in the neutron production target can be removed by the water cooling and the solid target can be used.

The heat load in the moderator is sufficiently small to be removed by the evaporation of liquid helium.
Pressurized circulation should be considered beyond $300$ W heat load when the beam power is increased.

The heat load in the UCN converter is about $10$ W.
The required cooling power is estimated in terms of the consumption of liquid helium at several converter temperatures.

\begin{itemize}
\item (1) $5$K\\
The helium consumption is 15 L/h and 360 L/day, which is calculated from the latent heat of liquid helium (20 kJ kg$^{-1}$).
A 1000 L vessel should be supplied every two days.
\item (2) $2$K\\
The helium consumption is 25 L/h and 600 L/day, which is estimated based on the experience of the Low Temperature Engineering Center of KEK: $15$ W cooling power at $2$ K.
A 1000 L vessel should be supplied every 1.3 days.
\item (3) $0.8$K\\
The helium consumption is 50-100 L/h and 1200-2400 L/day, which is a crude estimation since we have no experience.
One to three 1000 L vessel should be supplied every day.
\end{itemize}
In the present discussion, the UCN converter is assumed to be kept at $0.8$ K under the $15$ W heat load.

A new liquefier or refrigerator is recommended to satisfy the corresponding consumption of liquid helium~\footnote{
The practical way to employ a dedicated liquefier is to separate it from the UCN converter, which consumes the liquid helium, without connecting the liquefier and the UCN converter.
Otherwise, the national law on the safety of high pressure gas requires a special admission by the Minister of Industry of the Japanese Government.
}.
A new liquefier requires the increase of qualified manpowers for the safety and operation.
If we directly connect the refrigerator and the UCN converter, we may decrease the amount of qualified manpower for safety and operation spending more time and efforts to get the admission by the Government.

\subsection{Radiation Shield}
\label{sec:shield}
We require that the radiation dose rate on the shield surface is $1 \mu$Sv h$^{-1}$ or less, consistently with the safety regulation of the J-PARC MLF experimental hall~\footnote{
The design goal of the J-PARC Controlled Area Category 1 is $12.5 \mu$Sv h$^{-1}$.
This radiation dose rate is also considered and the result is shown in Table \ref{tab:shield}. 
}.
This safety regulation was determined for the safety and also for the reasonably low radiation background for physics measurements.

We determined the dimension of radiation shield by scaling the J-PARC MLF neutron source by the beam power.
\begin{figure}[htbp]
	\begin{center}
		\includegraphics[width=0.4\linewidth,clip]{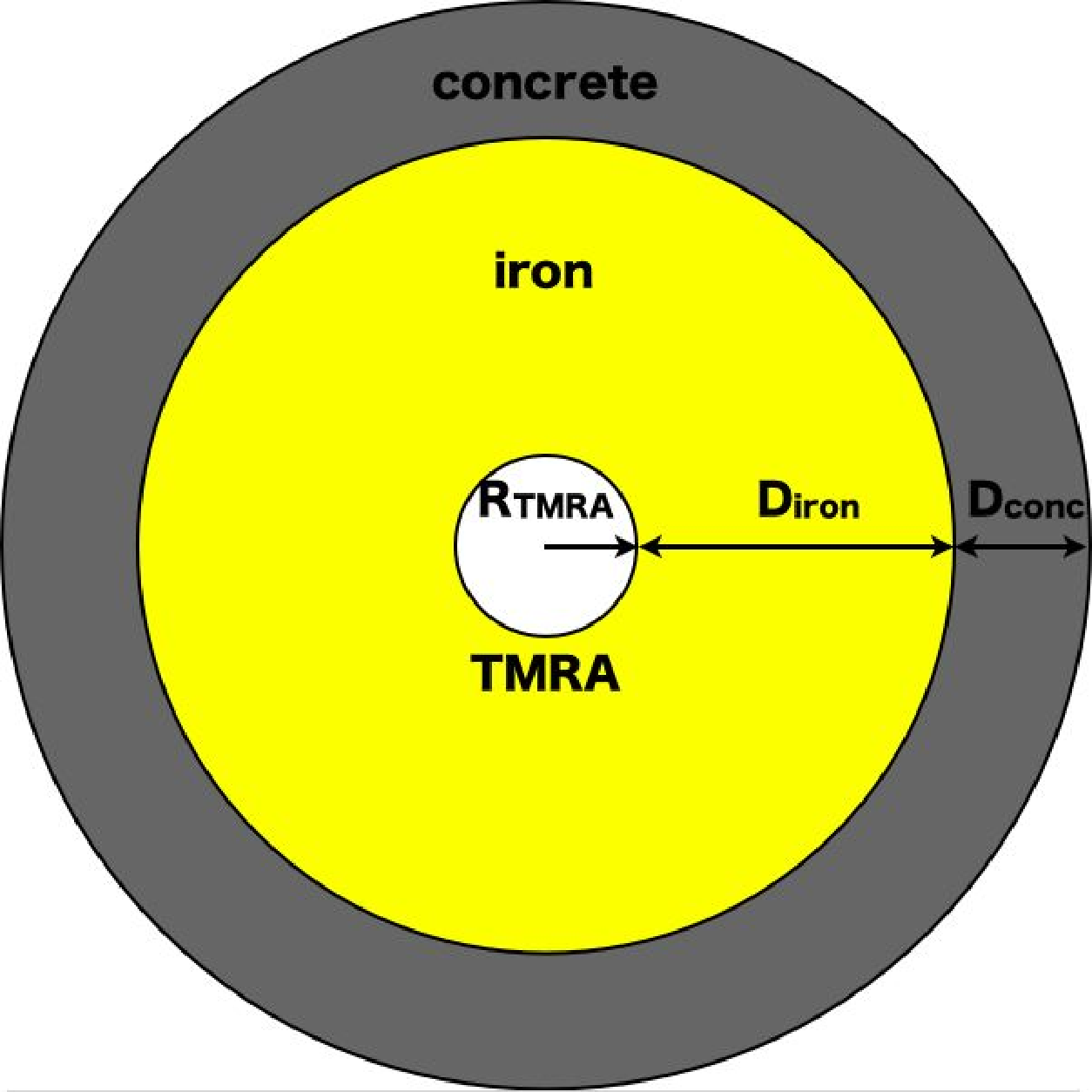}
	\end{center}
	\caption{
		Radiation shield structure used in the design.
	}
	\label{fig:shield}
\end{figure}

\begin{table}
	\begin{center}
	\begin{tabular}{|c|c|c|}
	\hline
	shield material & attenuation length & attenulation length \\
	 & ($3$ GeV proton) & ($400$ MeV proton) \\
	\hline
	iron & $0.244$ m & $0.177$ m \\
	concrete & $0.650$ m & $0.409$ m \\
	\hline
	\end{tabular}
	\caption{
	Attenuation length of iron and concrete~\cite{henko}. 
	}
	\label{tab:attenuation}
	\end{center}
\end{table}

\begin{table*}
	\begin{center}
	\begin{tabular}{|c||c||c|c||c|c|}
	\hline
	 & MLF & UCN & UCN & UCN & UCN \\
	\hline
	$E_p$ [MeV] & $3000$ & $400$ & $400$ & $400$ & $400$ \\
	Power [kW] & $1000$ & $20$ & $20$ & $200$ & $200$ \\
	surface dose rate & $1$ & $1$ & $12.5$ & $1$ & $12.5$ \\
	@[$\mu$Sv h$^{-1}$] & & & & & \\
	\hline
	core radius $R_{\mathrm{TMRA}}$ [m] (΁Տ) & $0.687$ & $0.69$ & $0.69$ & $0.69$ & $0.69$ \\
	iron thickness $D_{\mathrm{iron}}$ [m] & $4.113$ & $2.6$ & $2.4$ & $2.8$ & $2.6$ \\
	concrete thickness $D_{\mathrm{conc}}$ [m] & $3.200$ & $1.2$ & $0.7$ & $1.7$ & $1.2$ \\
	\hline
	outer diameter [m] & $16$ & $9.1$ & $6.2$ & $10.4$ & $8.9$ \\
	\hline
	\end{tabular}
	\caption{
	Estimation of required size of the radiation shield.
	}
	\label{tab:shield}
	\end{center}
\end{table*}

In the case of the UCN location candidate 1, the UCN source is neighboring to the tunnel for the future extension of the linear accelerator, in which there is no access during the beam operation.
If we assume that the radiation dose rate of $1$mSv h$^{-1}$ is acceptable during the operation in the accelerator tunnel, a $1.5$ m thick iron wall is sufficient for the beam power of $20$ kW.

\begin{figure*}[htbp]
	\begin{center}
		\includegraphics[width=0.95\linewidth,clip]{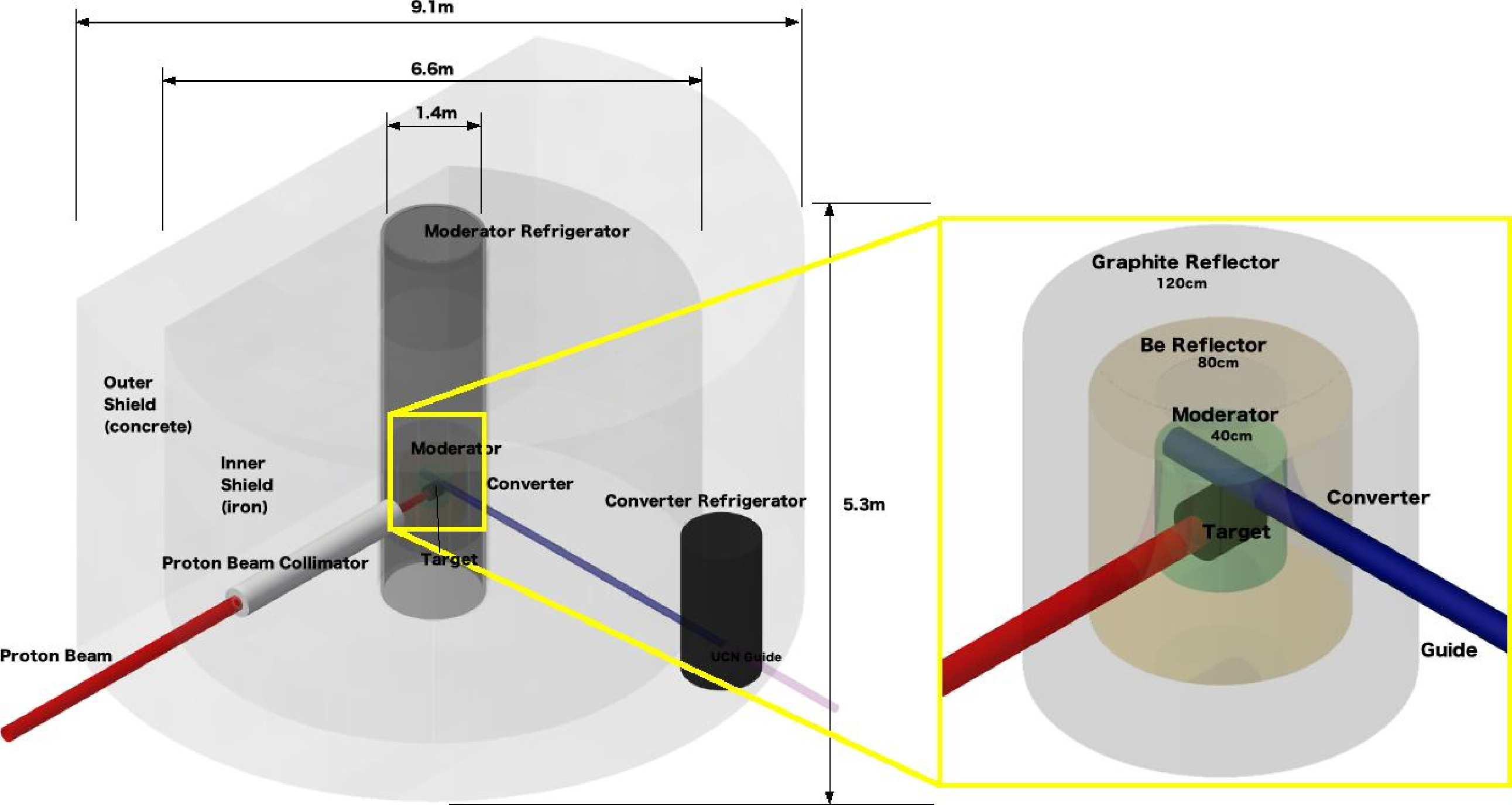}
	\end{center}
	\caption{
Configuration and outer dimension of the $20$ kW UCN source with the neutron production target, moderator, reflector, converter and radiation shield with the surface dose rate of $1 \mu$Sv/h.
In this example, the radiation shield on the left rear is decreased assuming that there is no access to that side during the operation.
	}
	\label{fig:tmras-20kW}
\end{figure*}

Under these assumptions, the minimum radiation shield is $117$m$^3$ ($916$ tons) of iron and $147$m$^3$ ($324$ tons) of concrete as shown in Fig.~\ref{fig:tmras-20kW}.
The rectangular hexahedron tangential to the minimum column is $150$m$^3$ ($1150$ ton) of iron and $169$m$^3$ ($373$ ton) of concrete.

\section{Construct Cost}
\begin{figure}[htbp]
	\begin{center}
		\includegraphics[width=0.90\linewidth,clip]{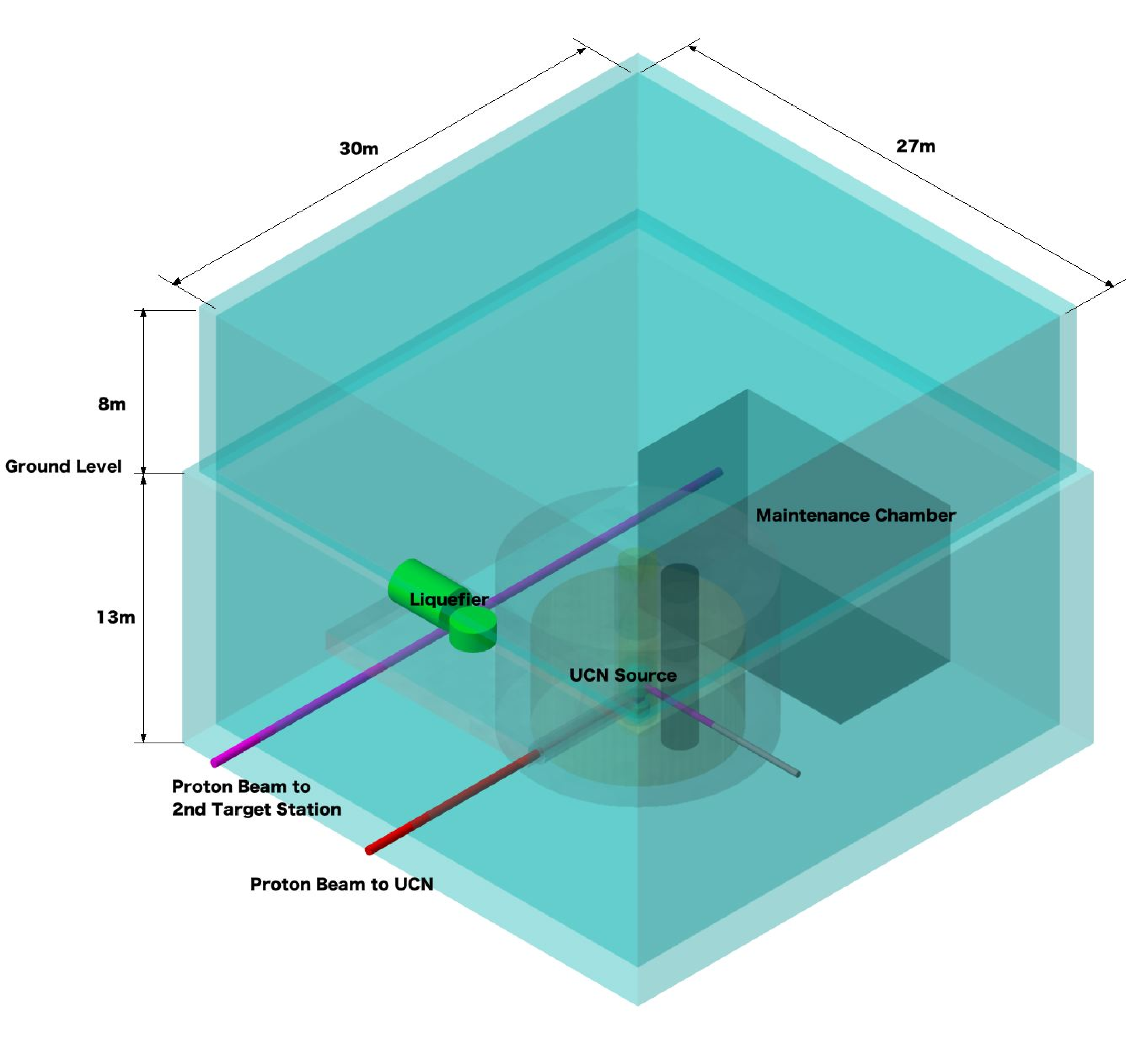}
	\end{center}
	\caption{
Building for the UCN source and the experiment for the case of UCN source location candidate 1.
The tunnel for the future extension of the linear accelerator is on the left rear of the UCN source.
	}
	\label{fig:ucn-building}
\end{figure}

\begin{figure}[htbp]
	\begin{center}
		\includegraphics[width=0.7\linewidth,clip]{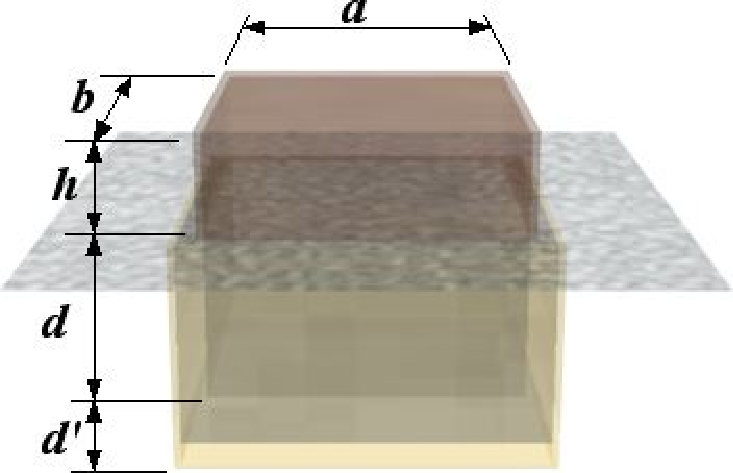}
	\end{center}
	\caption{
		Definition of the outer dimension of the building.
	}
	\label{fig:tateya}
\end{figure}

The dimension of the building to install the UCN source is shown in Fig.~\ref{fig:ucn-building}. 
The floor level of the building is $13$m below the ground level which is the accelerator building floor level.
The building dimension is $d^{\prime}=5$m $a=30$m, $b=27$m, $h=8$m, and $d=13$m as shown in Fig.~\ref{fig:tateya}.
The construction cost of the building is estimated as about $950$ million yen in rough figures. 

The rough cost estimate for the UCN source construction is listed in Table ~\ref{tab:budget}. 
\begin{table*}
	\begin{center}
	\begin{tabular}{|ll|c|l|}
	\hline
	& & (million yen) &  \\
	\hline
	building construction & & $950$ & \\
	\hline
	proton beam branch & pulse magnet & $50$ & \\
	and transport & DC magnets & $25$ & \\
	& quadrupoles and power supplies & $15$ & \\
	& control & $25$ & \\
	& linear part quadrupoles & $25$ & operated by \\
	& linear part quadrupole power supplies & $20$ & present LINAC team\\
	& steering magnet & $10$ & \\
	& steering magnet power supply & $10$ & \\
	& vacuum & $25$ & \\
	& beamline radiation shield & $15$ & \\
	\hline
	UCN source & production target & $45$ & 3 personnel \\
	& moderator & $100$ & (1 personnel all the time) \\
	& converter & $120$ & \\
	& reflector and radiation shield & $150$ & \\
	\hline
	heium liquefier & (120L/h) & $400$ & 1 qualified personnel \\
	& & & during operation \\
	\hline
	neutron transport &  & - & 1 technician \\
	\hline
	physics experiments &  & - & not involved in this report \\
	\hline
	total &  & $1,985$ & \\
	\hline
	\end{tabular}
	\caption{
	Crude cost estimation of the UCN source construction at the J-PARC.
	}
	\label{tab:budget}
	\end{center}
\end{table*}
The construction cost can be suppressed by the re-use of proton beam transport and radiation shield and by careful re-design of the building.
The period of the construction is estimated as $18$ months without any contingency.

\section{Near Fugure}
Details of the design and the physics program using the UCN source will be discussed in the fiscal year of 2009, which consequently expand the researchers' community.
Measurement techniques will be studied and developed using the cold beam line for the neutron optics at the BL05 of J-PARC MLF.
A new research collaboration framework will be promoted through above activities at nation-wide level and also at the international level.
The collaboration framework will pay efforts to obtain sufficient financial support.

\section{J-PARC UCN Taskforce Members}
\begin{tabular}{|ll|}
\hline
Tomokazu Aso & JAEA/J-PARC Neutron Source \\
Masatoshi Futakawa & JAEA/J-PARC Neutron Source \\
Tomiyoshi Haruyama & KEK IPNS \\
Kazuo Hasegawa & JAEA/J-PARC Accelerator \\
Yujiro Ikeda & JAEA/J-PARC MLF \\
Masanori Ikegami & KEK Accelerator \\
Yukihide Kamiya & KEK Accelerator \\
Takashi Kato & JAEA/J-PARC Neutron Source \\
Nobuhiro Kimura & KEK Cryogenics Science Center \\
Yoshiaki Kiyanagi & Hokkaido University \\
Yasuo Maekawa & JAEA/J-PARC Neutron Source \\
Yasuhiro Masuda & KEK IPNS \\
Taichi Miura & KEK Radiation Science Center \\
Masaharu Numajiri & KEK Radiation Science Center \\
Toru Ogitsu & KEK Cryogenics Science Center \\
Nobuo Ouchi & JAEA/J-PARC Accelerator \\
Kotaro Sato & KEK Accelerator \\
Hirohiko Shimizu & KEK IMSS Neutron Division \\
Eiichi Takasaki & KEK Accelerator \\
Nobuyuki Takenaka & Kobe University \\
Akira Yamamoto & KEK Cryogenics Science Center \\
Satoru Yamashita & ICEPP, University of Tokyo \\
\hline
\end{tabular}

\end{document}